\documentclass[fleqn,usenatbib]{mnras}

\usepackage{newtxtext,newtxmath}

\usepackage[T1]{fontenc}

\DeclareRobustCommand{\VAN}[3]{#2}
\let\VANthebibliography\thebibliography
\def\thebibliography{\DeclareRobustCommand{\VAN}[3]{##3}\VANthebibliography}

\usepackage{graphicx}	
\usepackage{amsmath}	
\usepackage{bm}

\usepackage{color}
\usepackage[normalem]{ulem}

\title[Fermi acceleration at oblique shocks]{Spectral curvature and breaks from Fermi acceleration at oblique shocks}

\author[A. Shirin T et al.]{
Asma Shirin T,$^{1,2}$
Brian Reville,$^{1}$\thanks{E-mail: brian.reville@mpi-hd.mpg.de}
Nils W. Schween,$^{1}$ Florian Schulze,$^{1}$ and John G. Kirk$^{1}$
\\
$^{1}$Max-Planck-Institut f\"ur Kernphysik, Saupfercheckweg 1, 69117 Heidelberg, Germany\\
$^{2}$Indian Institute of Science Education and Research, Tirupati, India
}

\date{Accepted Nov 2025. Received 2025}

\pubyear{\the\year{}}

\begin{document}
\label{firstpage}
\pagerange{\pageref{firstpage}--\pageref{lastpage}}

\maketitle

\begin{abstract}
 A major attraction of diffusive shock acceleration is the prediction of
power-law spectra for energetic particle distributions. However, this property
is not fundamental to the theory. We demonstrate that for planar shocks with
an oblique magnetic field the generation of power-law spectra critically
requires the particles' scattering rate to be both directly proportional to
their gyro radius ({\em Bohm scaling}) and spatially uniform. Non-Bohm scaling
results
in curved spectra at oblique shocks, while abrupt changes in the spatial profile of the scattering upstream introduces spectral breaks.
  Using the publicly available code \texttt{Sapphire++}, we numerically explore the magnitude of these effects, which are particularly pronounced at fast shocks, as expected in active galactic nuclei and microquasar jets, or young supernova remnants.
\end{abstract}




\section{Introduction}
Fermi acceleration at astrophysical shocks is commonly invoked to interpret the emission from sources of non-thermal radiation.
The process plays a central role in theories of cosmic-ray origins.
The non-relativistic test-particle derivation \citep{AxfordLeerSkadron, Krymskii, BlandfordOstriker, Bell78I}, commonly referred to as diffusive shock acceleration (DSA), provided the first reliable mechanism for production of power-law particle spectra, $f\propto p^{-q}$, extending over a large range in momenta.
Standard DSA theory has the appealing prediction that the particles' spectral index $q$ depends on only a single parameter: the compression ratio of the shock.
The theory was later generalised to relativistic shocks \cite[e.g.][]{Peacock, KirkSchneider87, HeavensDrury}, where despite additional complexity, a universal power-law index $q\approx4.2$ is found in the ultra-relativistic weakly-magnetised limit  \cite[e.g.][]{KirkGuthmann,Achterberg,Sironi,KirkReville23}.

On the other hand, many astrophysical sources reveal intrinsically curved particle spectra, or broken power laws \cite[e.g.][]{Slaneetal, Stawarzetal,2017ApJ...843..147M, Ajello2020,2024A&A...685A..96H}. The long-standing presumption that shock acceleration produces exclusively power-law spectra has prompted consideration of alternative acceleration mechanisms, such as stochastic acceleration, to explain such observations \cite[see for example][and references therein]{Kapanadze}.

In this letter we demonstrate that curved spectra and spectral breaks are naturally produced by Fermi acceleration
at shocks when the fundamental assumption that the angular distribution of
accelerated particles is everywhere almost isotropic, is broken by the influence of an ambient, large-scale magnetic field. 
This happens when two widely adopted simplifying assumptions are relaxed.
The first is the restriction to parallel shocks, where the average ambient magnetic field ($\mathbf{B}_{\rm up}$) is aligned with the shock normal ($\mathbf{n}$), and consequently,
affects neither the probability for a particle to undergo multiple shock crossings nor the energy boost it receives each time this occurs.
In the more general case of misalignment (\lq\lq oblique shocks\rq\rq), the magnetic field does play a role, leading to deviations from the DSA predictions
that become more pronounced for fast shocks with speeds $u_{\rm sh}> 0.01c$ and moderate obliquity, as shown, for example, by \citet{Bell2011} and \citet{Takamoto}.
However, obliquity alone is not sufficient to cause a deviation from a scale-free, power-law solution. For this, the second
simplifying assumption
adopted in these papers, namely that the scattering mechanism obeys Bohm energy-scaling, must be relaxed.
Bohm scaling assumes that the scattering frequency is proportional to the particle gyrofrequency, making it impossible to identify
a characteristic particle energy at a planar shock in an otherwise uniform, infinite medium.
When this condition is relaxed, or if the scattering rate has a significant spatial dependence near the shock, a characteristic particle energy emerges, leading to spectra that depart from a single power-law.

The problem of particle injection into the DSA mechanism at oblique shocks
of high Alfv\'en Mach number is relevant in many astrophysical
contexts, such as young supernova remnants and pulsar wind nebulae, as well
as in the collimated outflows observed in microquasars and active galactic nuclei.
Recent kinetic simulations of shock acceleration \cite[e.g.][]{Xu, Kumar, vanMarle, Orusa}
have reignited interest in this topic, extending previous theoretical work
\cite[see for example][]{Ostrowski88,KirkHeavens,Baring,Zank06}. 

Adopting an isotropic, small-angle scattering model,
\citet{Bell2011} demonstrated that the steady-state non-thermal particle spectrum produced in 1D shock simulations is sensitive to the magnetic field orientation, the shock velocity, and the scattering rate.
Their numerical approach involved an expansion of the particle phase-space density $f(x,\mathbf{p})$, specifically its momentum dependence in terms of spherical harmonics.
This transforms the Vlasov-Fokker-Planck (VFP) equation (which depends on three momentum coordinates, see Appendix \ref{App:VFP}) into a system of equations for the expansion coefficients, which are functions only of $p$ (and $x$).
In all of their simulations, Bohm scaling was used for the scattering rate, i.e., $\nu = \eta \omega_g$, where $\omega_{\rm g} = |eB|/\gamma mc$ is the relativistic gyrofrequency and $\eta$ was chosen to be constant and uniform.
With Bohm scaling, the equations are self-similar for relativistic particles and admit power-law solutions of the form $f \propto p^{-q}$.
The problem is then reduced to finding the correct value for $q$ at an arbitrary momentum $p$.
An expression for $q$ may be derived by integrating the zeroth order equation across the shock, which to lowest order in $u/c$ reads \citep{Bell2011}:
\begin{equation}
  \label{eq:spectral_index}
  q_0 =\left.-\frac{\partial \ln f_{0}}{\partial \ln p}\right|_{x=0}= \bar{q} + \frac{3}{r-1}\left(\rho_{\rm f}-1\right)\, ,
\end{equation}
where $\bar{q} = {3r}/(r-1)$ is the test-particle solution for non-relativistic parallel shocks in the diffusion approximation \cite[e.g.][]{Drury83}.
Here, $r$ is the shock-compression ratio, while
\[\rho_{\rm f}(p) \equiv \frac{f_{0}(p,x=\infty)}{f_{0}(p,{x=0})}\]
is the ratio of the isotropic particle density at the shock position to its asymptotic value far downstream, where the isotropic particle density is $f_0(p,x)=\frac{1}{4 \pi}\int f(\mathbf{p},x) \, d\Omega$.
In general, it is necessary to find $\rho_{\rm f}$ numerically.

In this work, we revisit the acceleration of test particles at oblique shocks, extending the results of \citet{Bell2011}.
We consider steady-state solutions to the Vlasov-Fokker-Planck equation in one spatial dimension, using the \texttt{Sapphire++} code, which accurately captures high-order anisotropies in momentum space, and a large dynamic range in momentum.
In the next section, we describe our physical setup, and introduce the key features of the numerical approach.
Results are presented in section \ref{sec:results}.
We conclude in section \ref{sec:conc}.

\section{Methods}

To proceed, we make use of the \texttt{Sapphire++} code \citep{Schween2025}.
We implement a new algorithm that calculates the steady-state solution to the VFP equation for a single species of mass $m$ and charge $e$, in a prescribed shock profile (see Appendix \ref{App:VFP} for details of the numerical method).
Particle scattering on magnetic field perturbations is modelled assuming small-angle isotropic collisions, with scattering rate $\nu(p)$.
We consider a stationary flow profile
\begin{equation}
  \label{eq:velocity-profile}
  \mathbf{u} = \frac{u_{\rm sh}}{2r}\left[(r+1) - (r-1) \tanh\left(\frac{x}{L_{s}}\right)\right] \hat{\mathbf{x}}\,.
\end{equation}
representing a planar shock of thickness $\approx L_s$ in one spatial dimension, centred on $x=0$.
The finite width is required for numerical reasons, but provided $L_s$ is much less than the gyroradius $r_{\rm g}$ of the lowest energy particles considered, it has a negligible effect on the solution \cite[e.g.][]{1982MNRAS.198..833D}.
Without loss of generality, we orient the average upstream magnetic field in the $x-z$ plane.
For simplicity, we take $r=4$ as expected for a strong shock in a non-relativistic monatomic gas with a high Alfv\'en Mach number, $M_{\rm A} \gg 1$.
We neglect the small bulk velocity that appears in the $z$ direction $u_z \sim u_{\rm sh}/M_{\rm A}^2 \ll u_x$, downstream of oblique shocks \cite[e.g.][eq. 13]{Decker}.
Thus, the steady-state Maxwell-Faraday equation in the ideal-MHD approximation enforces $u_x B_z$ to be constant, and hence $B_z$ also varies smoothly across the shock.

All simulations are performed using a 2D grid in $x$-$\ln p$ space, with stretched spatial coordinates that accurately resolve the shock transition width $L_s = 0.1 r_{\rm g,0}$, where $r_{g,0} = mc^2/eB_0$ is a reference gyroscale, with $B_0$ the magnitude of the upstream magnetic field.
The minimum step size close to $x=0$ is $\Delta x_{\rm min} = L_s/10$.
Free-escape boundary conditions are implemented at the upstream boundary, and continuous boundary conditions are used at the downstream boundary.
To accommodate a large momentum range, the grid spacing in the \(x\)-direction is increased exponentially with distance from the shock, covering $\approx6$ times the maximum upstream diffusion length, to avoid additional spectral features due to free-escape.
Continuous boundary conditions are applied at the end points of the momentum grid.
Particles are continuously injected with a narrow Gaussian distribution centred on $p_{\rm inj} = 2 mc$.
The gyroradius at $p_{\rm inj}$ in the upstream field is therefore $20 L_s$. 

As a validation of the new code, a comparison of our solutions with published results of \citet{Bell2011} is presented in Appendix \ref{App:BellComp}.
We also checked that the finite shock thickness had negligible impact on the solution by comparing the reconstructed phase-space distributions immediately up and downstream in a common reference frame.
Liouville's theorem implies that these should be identical in the absence of collisions within the shock transition.
In our simulations, they agree to the accuracy of the code, i.e., to order $(u/c)^2$.
We note that $f_0$ undergoes only a small change across the shock transition, but can asymptote to a value where $\rho_{\rm f}$ differs noticeably from unity.
The asymptotic state is reached at distances downstream on the order of one gyroradius.

\begin{figure}
  \centering
  \includegraphics[width=0.95\linewidth]{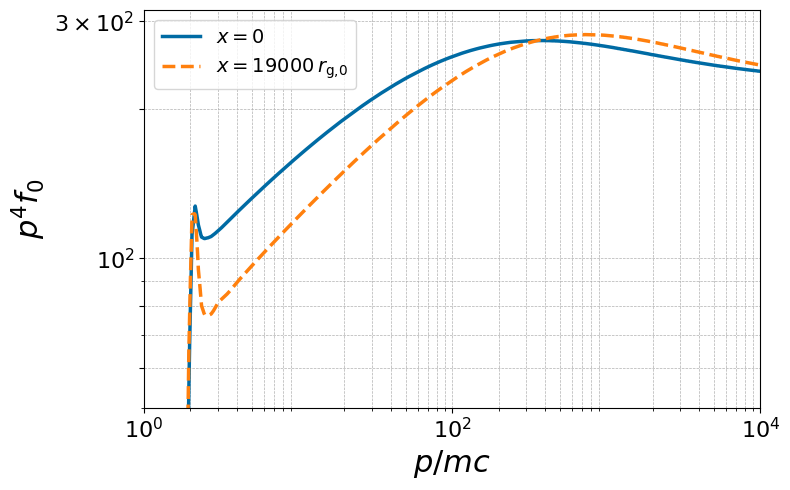}
  \caption{Particle spectrum at the shock and far downstream, accelerated at a shock with magnetic obliquity $\theta_n = 60^\circ$ and $u_{\rm sh} = 0.1 c$.
    The scattering rate follows Kraichnan scaling, i.e. $\nu/\omega_{\rm g}= \sqrt{p/p_{\rm Bohm}}$.
    The scattering rate meets the Bohm limit, i.e. $\nu = \omega_{\rm g}$, at $p_{\rm Bohm} = 10^4 mc$.
  }
  \label{fig:1}
\end{figure}

Such jumps are expected at oblique shocks \cite[see][and references therein]{Gieseler}.
In the scatter-free adiabatic approximation, the magnetic moment is assumed to be the same before and after interaction with a subluminal shock \cite[though not during, e.g.][]{Decker}.
Applying this ansatz to particles incident on the shock from far upstream, defines a critical incident pitch angle (w.r.t. the B-field) that separates transmitted from reflected particles.
All particles incident from far downstream are transmitted.
In this approximation, a jump in the density distribution, of order $u/c$, is expected across an oblique shock \citep{Gieseler}. In this work \cite[see also][]{Bell2011}, the gyromotion about the shock is well resolved, and conservation of the magnetic moment is not enforced.
This smears the jump over the scale of the particle's gyroradius, and is clearly visible in the spatial profiles (see Figures \ref{fig:5} and \ref{fig:6}).
We note that the above argument holds only for subluminal shocks, i.e. those for which the point of intersection between field lines and the shock surface do not exceed the speed of light.
For superluminal shocks, magnetised particles can never outrun the shock, and thus cannot be reflected.
While scattering will result in some cross-field diffusion that may allow repeated shock crossings, unless scattering is strong, particles are trapped to field lines and advected downstream.
This effectively enhances the escape probability which leads to steeper spectra.
In contrast, shock reflection at moderate obliquities reduces the escape probability, and hard spectra ($q<4$) can result.

\section{Results}
\label{sec:results}

Next we explore two new physical scenarios which introduce previously unidentified features in the non-thermal particle spectrum at oblique astrophysical shocks: (i) the impact of non-Bohm scaling and (ii) the introduction of an enhanced scattering zone in the immediate shock upstream.

\subsection*{(i) Non-Bohm scattering}

We consider the case of Iroshnikov–Kraichnan MHD turbulence, in which the momentum dependence of the scattering scales as $\nu \propto p^{-1/2}$ \cite[e.g.][]{2005ppa..book.....K}, although we maintain the isotropic scattering assumption.
Since $\omega_{\rm g} \propto p^{-1}$, one can define a critical momentum $p_{\rm Bohm}$ at which $\nu = \omega_{\rm g}$ and above which the scattering rate $\nu > \omega_{\rm g}$, i.e. sub-Bohm, a regime that is not physically well motivated.
We therefore ignore particles with $p>p_{\rm Bohm}$.

\begin{figure}
  \centering
  \includegraphics[width=\linewidth]{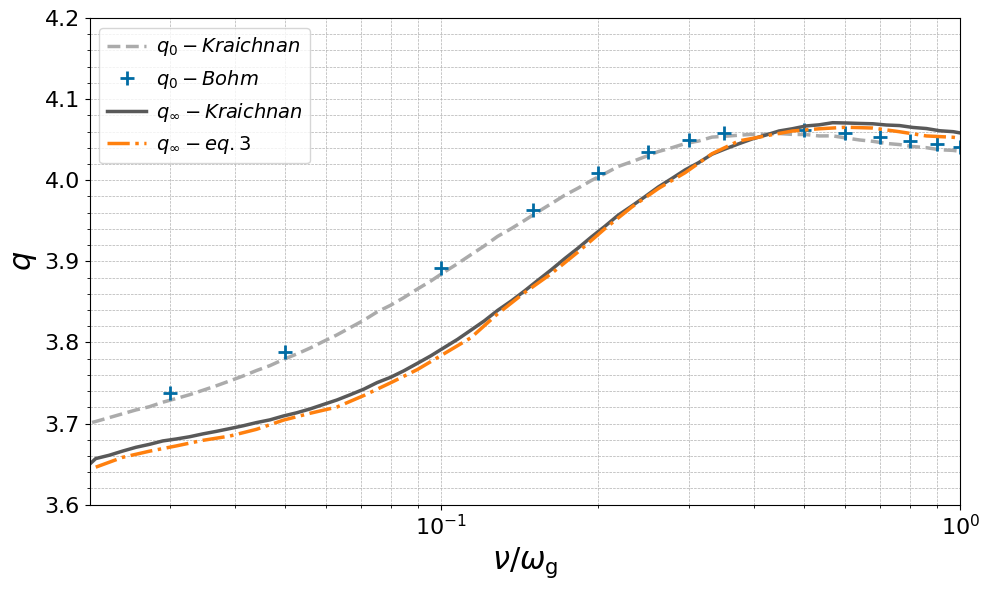}
  \caption{ Variation of the spectral index ($q = -{\partial \ln f}/{\partial \ln p}$) at $x=0$ ($q_0$) and far downstream, at $x=1.9\times10^4 r_{\rm g,0}$ ($q_\infty$), for the same simulation as in Fig.~\ref{fig:1}.
  For comparison with Bohm scaling simulations (Blue crosses), we use $\nu/\omega_{\rm g} = \sqrt{p/p_{\rm Bohm}}$ for the $x$-axis.
  The theoretical curve for $q_{\infty}$ line is calculated using the simulation result for $q_0$ with eq. \eqref{eq:q_infty}.
  }
  \label{fig:2}
\end{figure}

We focus on shocks with $u_{\rm sh}=0.1 c$, and for this example we restrict ourselves to $\cos \theta_n = 0.5$.
Particles are injected with a Gaussian spatial profile about $x=0$.
We choose $p_{\rm Bohm} = 10^4mc$, such that the non-thermal spectrum covers a range $0.02 < \nu/\omega_{\rm g} < 1$.
Fig.~\ref{fig:1} shows the resulting steady-state particle spectrum at the shock, and far downstream.
The solution does not follow a simple power-law distribution, but rather concave downwards, i.e. harder at low energies and softer at high energies.
This shape could in principle have been predicted from earlier simulations, since the spectrum was already known to be sensitive to the scattering rate \citep{Bell2011}.
In Fig.~\ref{fig:2}, we compare the spectral index, as a function of momentum, with a suite of simulations that use Bohm scaling with different (fixed) values of $\eta=\nu/\omega_{\rm g}$.
For direct comparison, we use the fact that for Kraichnan-like scattering, $\nu/\omega_{\rm g} = \sqrt{p/p_{\rm Bohm}}$, i.e. we plot the function $q\left(p=p_{\rm Bohm}\nu^2/\omega_{\rm g}^2\right)$.

Since emission in astrophysical systems is typically dominated by particles accumulated downstream, the characteristics of this region's particle spectrum are most relevant for comparison to observations.
From Fig.~\ref{fig:1}, we see that the spectrum far downstream of the shock differs from that at $x=0$.
This is an unavoidable consequence of the \(p\)-dependence of $q_0$.
As equation\,(\ref{eq:spectral_index}) holds irrespective of the scattering law, the \(p\)-dependence of \(q_{0}\) implies that \(\rho_{f}\) must also be \(p\)-dependent; \(\bar{q}\) and \(r\) are, by definition, independent of \(p\).
An expression for the asymptotic downstream spectrum may be found by differentiating equation\,(\ref{eq:spectral_index}) w.r.t \(p\) and rearranging, to yield:
\begin{equation}
  \label{eq:q_infty}
  q_{\infty} = \left.-\frac{\partial \ln f_{0}}{\partial \ln p}\right|_{x=\infty}
  = q_0 - \frac{(r-1)}{3+(q_0-\bar{q})(r-1) } \frac{\partial  q_0}{\partial \ln p} \,.
\end{equation}
This shows that the slope of the spectrum far downstream \(q_{\infty}\), i.e. its curvature, differs from the slope of the spectrum at the shock \(q_{0}\), namely by the second term in equation\,(\ref{eq:q_infty}).
In Fig.~\ref{fig:2}, we plot \(q_{\infty}\) once using equation\,(\ref{eq:q_infty}) and a second time extracting it directly from the far downstream spectrum shown in Fig.~\ref{fig:1}.
We note that the \(q_{0}\) values in equation\,(\ref{eq:q_infty}) are also computed using the particle spectrum at the shock as shown Fig.~\ref{fig:1}.
There is good agreement between the two curves.

\subsection*{(ii) Spatial dependent scattering}

We next consider the impact of spatially dependent scattering in the shock precursor.
This is motivated by numerical shock simulations that show a build up of self-excited magnetic fields in the shock precursor, peaking close to the shock \citep{Bell2013, Reville2013, Downes,2014ApJ...794...46C}.
As an illustrative example, we take the case of a mildly superluminal shock with $\theta_n = 85^\circ$ and $u_{\rm sh}=0.1 c$.
Such conditions might occur in the internal shocks of AGN jets
\cite[e.g.][]{2025arXiv250309180S}.

We construct an artificial scattering profile, imposing Bohm scaling, but changing the scattering behaviour as a function of position.
Evaluating the transmission of even weak perturbations through a strong shock is not trivial \cite[e.g.][]{1982PhFl...25..748Z}, and the nature of scattering downstream is poorly constrained.
We consider therefore two possibilities,
where either $\eta$ or $\nu$ undergo a jump at the shock.
This is achieved with the spatial profile
\begin{equation*}
  \zeta(x) = \frac{\zeta_{\rm 1}+\zeta_{\rm 2}}{2} - \frac{\zeta_{\rm 1}-\zeta_{\rm 2}}{2} \tanh\left(\frac{x+x_{\rm pre}}{L}\right)\,,
\end{equation*}
i.e. $\zeta = \zeta_{\rm 1}$ far upstream,
and transitions to $\zeta_{\rm 2} (>\zeta_{\rm 1})$ about $x=-x_{\rm pre}$,
persisting with the same value into the downstream.
We consider two cases
(a) $\eta = \zeta$,
and (b) $\nu = \omega_{\rm g,up} \zeta$, giving, in each case, three different scattering regimes: far upstream, precursor and downstream.

We can interpret the physical picture presented in the two cases as follows.
For non-parallel shocks, $\omega_{\rm g}$ undergoes a jump across the shock.
Taking case (a) $\eta = \zeta$,
the scattering rate $\nu$ must also jump across the shock,
giving three different scattering regimes,
the far upstream $\nu_{\rm up} = \zeta_{\rm 1} \omega_{\rm g,up}$,
in the precursor $\nu_{\rm pre} = \zeta_{\rm 2} \omega_{\rm g,up}$,
and downstream $\nu_{\rm down} = \zeta_{\rm 2} \omega_{\rm g,down}$.
This leads to an enhanced scattering rate in the downstream,
$\nu_{\rm down} > \nu_{\rm pre} > \nu_{\rm up}$.
In case (b) the scattering rate $\nu = \omega_{\mathrm{g, up}} \zeta$ does not jump across the shock and
the three regimes for $\eta$ are,
$\eta_{\rm up} = \zeta_{\rm 1}$,
$\eta_{\rm pre} = \zeta_{\rm 2}$,
and $\eta_{\rm down} = (\omega_{\rm g,up}/\omega_{\rm g,down}) \zeta_{\rm 2}$.
In this case $\eta$ decreases on crossing the shock,
$\eta_{\rm down} < \eta_{\rm pre}$,
implying that the shock compressed average field plays the more important role.
These simple test cases are designed to illustrate the sensitivity of the spectra to the scattering details.

\begin{figure}
  \centering
  \includegraphics[width=1\linewidth]{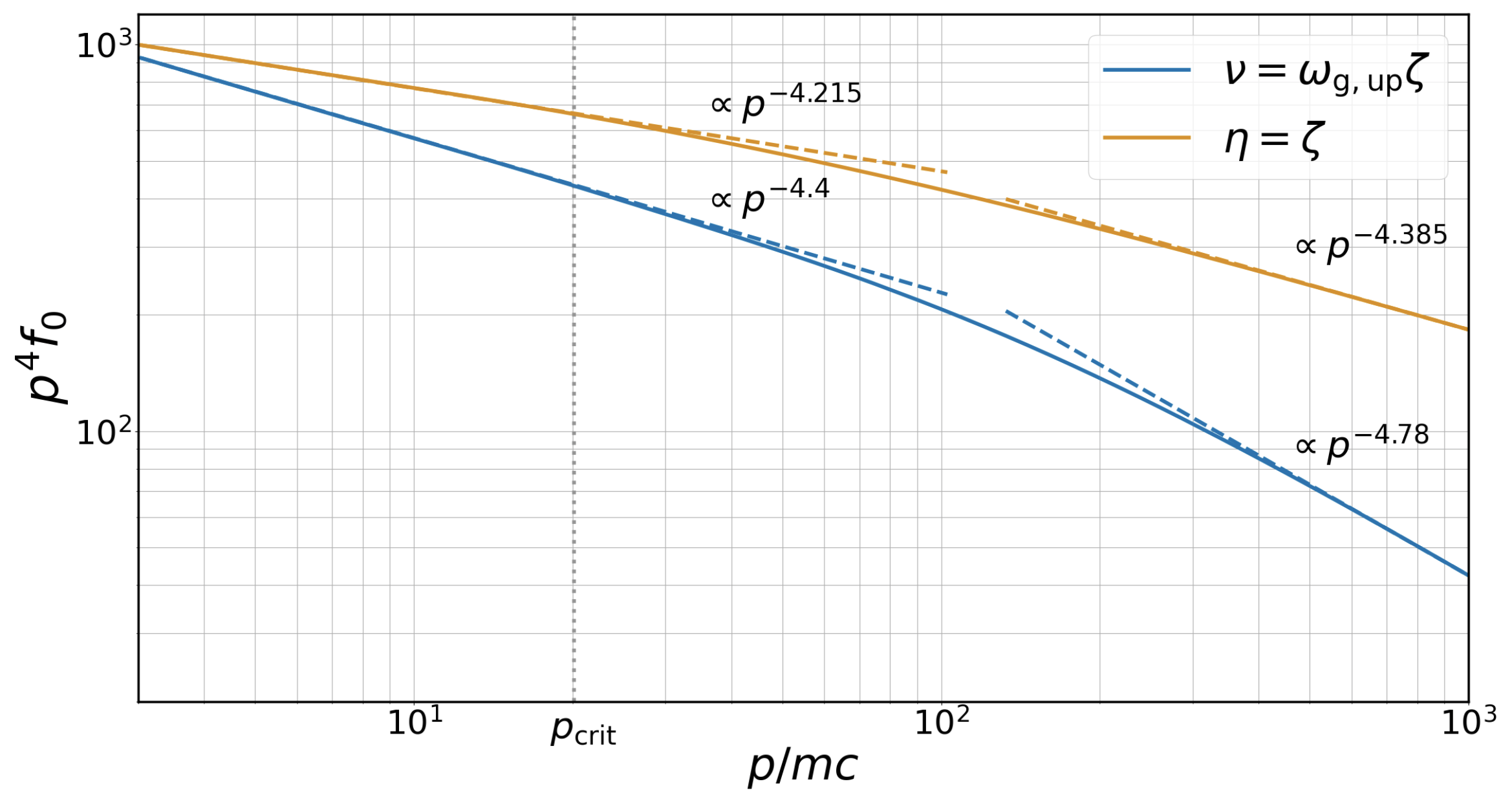}
  \caption{
    A comparison of a solution for a shock with $u_{\rm sh} = 0.1$ and $\theta_n=85^\circ$
    in the presence of a strong scattering precursor of thickness $20 r_{\rm g,0}$ upstream of the shock.
    In both cases $\nu_{\rm up} = 0.03 \omega_{\rm g,up}$ far upstream,
    and $\nu_{\rm pre} = 0.3 \omega_{\rm g,up}$ in the precursor.
    For the $\nu$ fixed curve, $\nu$ takes the same value in the downstream
    ($\nu_{\rm down} = \nu_{\rm pre}$),
    while for $\eta$ fixed, $\nu$ jumps by a factor
    $\nu_{\rm down} = (\omega_{\rm g,down}/\omega_{\rm g,up}) \nu_{\rm pre}$.
  }
  \label{fig:3}
\end{figure}

We set $\zeta_{\rm 1}=0.03$ and $\zeta_{\rm 2}=0.3$.
Note that for case (b) this relates to a downstream value $\eta_{\rm down} \approx 0.075$.
We set $x_{\rm pre} = 20 r_{\rm g,0}$ and a transition length scale $L = x_{\rm pre}/5$.
Thus low energy particles ($p<20 mc$) will see strong scattering both upstream and downstream
while higher energy particles sample the lower scattering rate in the far upstream.
The results are plotted in Fig.~\ref{fig:3}.

Both curves show a break at a momentum corresponding to the diffusion length matching the buffer zone width;
i.e., defining the upstream diffusion length \cite[e.g.][eq. 8]{Reville2013}
\begin{equation*}
  L_{\rm diff} = \frac{r_{\rm g}}{3 \eta_{\rm pre}} \frac{c}{u_{\rm sh}}\frac{\eta_{\rm pre}^2 + \cos^2 \theta_n}{\eta_{\rm pre}^2 +1}
\end{equation*}
the onset of the break occurs approximately where the momentum satisfies $L_{\rm diff}(p_{\rm crit}) = |x_{\rm pre}|$.
The spectral indices above and below the break match that for constant scattering rates
using $\eta_{\rm up}$ or $\eta_{\rm pre}$ as appropriate in upstream half-plane,
and $\eta_{\rm down}$ in the downstream.
The magnitude of the break when $\nu = \omega_{\mathrm{g, up} }\zeta$ is larger than when $\eta = \zeta$.
This is expected, since for superluminal shocks with low $\eta$ in the downstream,
particles are more effectively tied to field lines and the return probability is reduced.

The spectral shape depends on the magnetic field angle: $\theta_n = 85^\circ$ yields a spectral downturn, while $\theta_n = 60^\circ$ produces an upturn.
The sharpness of this spectral feature (whether a distinct break or more gradual curvature) directly reflects the sharpness of the change in upstream scattering.
A more gradual decay in the upstream scattering rate would result in a more concave spectrum.
This concavity can be concave-down (similar to Fig \ref{fig:1}) for superluminal shocks, or concave-up  \cite[resembling non-linear DSA predictions, e.g][]{2001RPPh...64..429M} for moderately oblique shocks.

\section{Discussion}
\label{sec:conc}

In this work, we demonstrated that Fermi acceleration at shocks, contrary to common perception, can exhibit a broad range of spectral features.
This has implications for the interpretation of observational data.
We focused here on a narrow range of the available parameter space to highlight the essential ideas and outline the key underlying physical mechanisms.
One could for example consider the combined action of spatial and energy dependent scattering rates, leading to additional features.
These effects are inevitable in non-linear DSA scenarios, where the cosmic-ray feedback affects both the magnetic field and global flow profile.

Although our straightforward 1D test-particle studies limit our ability to examine relevant multi-dimensional effects, it is still essential to avoid introducing any energy scale into the problem to achieve a pure power-law solution. This necessitates fine-tuning, which we believe is unlikely to occur naturally.

Examples to reproduce the results presented here are provided in the \texttt{Sapphire++} code repository.
Different choices for $u_{\rm sh},\ \theta_{n}$ and $\nu(p)$ could be used to compare with observational data.
This in turn will improve the interpretive power of non-thermal observations of many astrophysical sources, probing extreme environments such as blazar jets or other systems with fast shocks.

\section*{Acknowledgements}

The authors thank T. Bell for comments on an early draft, and F. Dann for additional input. We also express our gratitude to the referee.
BR thanks P. Duffy and S. Boula for feedback on the manuscript.
We acknowledge the staff at MPIK that maintain the computing cluster which was essential to carrying out this work.

\section*{Data Availability}

All results presented here are produced using the open-source \texttt{Sapphire++} code.
Example scripts to reproduce the results are available to download at \url{https://sapphirepp.org}.




\bibliographystyle{mnras}
\bibliography{references} 



\appendix

\section{Vlasov-Fokker-Planck equation}
\label{App:VFP}

The VFP equation, to order $u/c$, reads \citep{Skilling1975}:
\begin{eqnarray}
  && \left(1+\frac{\mathbf{u}\cdot\mathbf{v}}{c^2}\right)\frac{\partial f}{\partial t}+ \left(\mathbf{u} + \mathbf{v}\right) \cdot \nabla_{x} f
  + e \left(\frac{\mathbf{v}}{c}\times\mathbf{B}\right)  \cdot \nabla_{p}f  \nonumber\\
  &&\enspace \enspace -\left[ \gamma m\frac{d\mathbf{u}}{dt}+ (\mathbf{p}\cdot \nabla_{x})\mathbf{u}\right]\cdot \nabla_{p}f
  = \frac{\nu}{2} \Delta_{\theta\varphi}f + S \,.
  \label{eq:VFP}
\end{eqnarray}
Equation (\ref{eq:VFP}) uses mixed coordinates wherein time, space and $\mathbf{u}$ are measured in a fixed laboratory frame, while the fields, particle momenta ($\mathbf{p}= \gamma m \mathbf{v}$), and the momentum dependent scattering rate, $\nu(p)$ are measured in the local fluid frame.
It is assumed that ideal MHD applies, such that the electric fields vanishes in the fluid frame. $S(\mathbf{x},\mathbf{p})$ represents injection.

According to quasi-linear theory calculations of gyro-resonant scattering of relativistic particles on statistically isotropic magnetic fields fluctuations with power spectrum $\delta B^2 \propto k^{-\sigma}$,  the pitch angle scattering satisfies
\cite[e.g.][]{Jokipii71}
\begin{eqnarray}
D_{\mu\mu}=\frac{\langle \Delta \mu ^2\rangle}{2 \Delta t} \propto 
p^{\sigma-2}(1-\mu^2)|\mu|^{\sigma-1}\enspace.
\end{eqnarray}
For $\sigma > 1$, this vanishes at $\mu=0$, resulting in the well-known problem that particles are then confined in forwards and backwards moving hemispheres \cite{1973Ap&SS..25..471V}
thereby preventing their participation in DSA.

Test-particle simulations in synthetic turbulence, however, often find $D_{\mu\mu}/(1-\mu^2)$ is in practise a reasonably flat function of $\mu$ \cite[see][and references therein]{2024ApJ...977..174V}. For simplicity, we have therefore assumed scattering is isotropic (i.e. we neglect the additional $|\mu|$ dependence), but retain the momentum scaling. This is equivalent to expressing the scattering as the angular part of the Laplacian operator in spherical momentum coordinates, 
$\Delta_{\theta\varphi}$.

To find solutions, we make use of the open-source \texttt{Sapphire++} code\footnote{See \url{https://sapphirepp.org}}.
Details on its numerical implementation can be found in \citet{Schween2025}. \texttt{Sapphire++} enables coverage of a large dynamic range in 3D-momentum space by using a truncated real spherical harmonic expansion of \(f\), i.e.
\begin{align}
  \label{eq:real-spherical-harmonic-expansion}
  f(t,\mathbf{x}, p, \theta, \varphi) & = \sum^{l_{\mathrm{max}}}_{l = 0}\sum^{1}_{s = 0} \sum^{l}_{m = s} f_{lms}(t, \mathbf{x}, p) Y_{lms}(\theta, \varphi) \,.
\end{align}
This expansion is used to transform equation (\ref{eq:VFP}) to a system of coupled partial differential equations \cite[see][]{Schween2024}.
In the main text, we use the isotropic phase space density $f_0 = f_{000}Y_{000}$.

To solve equation  (\ref{eq:VFP}), we plug into it the spherical harmonic expansion of \(f\) given in eq.~\eqref{eq:real-spherical-harmonic-expansion} and derive a system of partial differential equations (PDEs) for the expansion coefficients \(f_{lms}\) using an operator based approach \citep{Schween2024}. In steady state, this leads to equations of the form
\begin{equation}
  \label{eq:system-pdes-expansion-coefficients}
  \bm{\beta}_{x}\frac{\partial \mathbf{f}}{\partial x}
  +   \bm{\beta}_{y}\frac{\partial \mathbf{f}}{\partial y}
  +   \bm{\beta}_{z}\frac{\partial \mathbf{f}}{\partial z}
  + \bm{\beta}_{p} \frac{\partial \mathbf{f}}{\partial p}
  + \mathbf{R} \mathbf{f} = \mathbf{S} \,,
\end{equation}
where \(\mathbf{f}\) is a vector whose components are the expansion coefficients \(f_{lms}\).
We call the matrices \(\bm{\beta}_{a}\) with \(a \in \{x,y,z,p\}\) the advection matrices and \(\mathbf{R}\) the reaction matrix.
The elements of the vector \(\mathbf{S}\) are the coefficients of the spherical harmonic decomposition of the source \(S\) i.e.
\begin{equation}
  \label{eq:source-term}
  S(x,p) = \frac{Q_0}{4\pi p^2} \frac{\exp\left[-\left(\frac{(x - x_{\text{inj}})^2}{2 \sigma^2_x} + \frac{(p -p_\text{inj})^2}{2\sigma^2_p}\right)\right]}{2\pi \sigma_x \sigma_p} \,,
\end{equation}
where \(Q_0\) determines the rate at which particles are injected per unit area.
We note that the spherical harmonic decomposition of \(S\) is
\(
S_{000}(x,p) = \int Y_{000} S \, \mathrm{d}\Omega = \sqrt{4\pi} S \,.
\)
For explicit expressions of \(\bm{\beta}_{a}\) and \(\mathbf{R}\), we refer the reader to \citet{Schween2025}.

In \texttt{Sapphire++} this system of PDEs is solved with the discontinuous Galerkin (dG) method.
We note that \texttt{Sapphire++} builds on the finite element (FE) library deal.II \citep{ dealii2019design,dealii2024}.
The idea behind the FE approach may be \emph{sketched} as follows: the approximate solution of the system of PDEs is written as a linear combination of, for example, polynomials that are defined on cells that result from decomposing the domain of \(f\) into a computational grid.
For the case at hand, we write
\begin{equation}
  \label{eq:approximate-solution}
  \mathbf{f}_{h} = \sum^{N - 1}_{i = 0} \zeta_{j} \bm{\phi}_{j}(\mathbf{x}, p) \,,
\end{equation}
where \(N\) is the total number of polynomials \(\bm{\phi}_{j}\), commonly called the number of degrees of freedom.
The subscript \(h\) indicates that the polynomials are defined on cells with size \(h\).
The \(\bm{\phi}_{j}\) are themselves vectors with their components each being polynomials.
The ansatz \(\mathbf{f}_{h}\) is plugged into the eq.~\eqref{eq:system-pdes-expansion-coefficients}.
To extract an algebraic system of equations for the coefficients \(\zeta_{j}\) of the linear combination.
We take the scalar product of the result with all \(\bm{\phi}_{i}\) where \(i \in {0, \dots, N - 1}\) and, subsequently integrate over \(\mathrm{d}x^{3} \, \mathrm{d}p\).
This yields $\mathbf{D}\bm{\zeta} = \mathbf{h}$.
Explicit expressions for \(\mathbf{D}\) and \(\mathbf{h}\) can be found in \cite[][eq.(37)]{Schween2025}. In \texttt{Sapphire++} this system of equations is solved iteratively using a preconditioned GMRES method, that is implemented in the PETSc library \citep{petsc-efficient, petsc-user-ref}.

\section{Comparison to Bell et al. 2011}
\label{App:BellComp}

\begin{figure}
  \centering
  \includegraphics[width=0.9\linewidth]{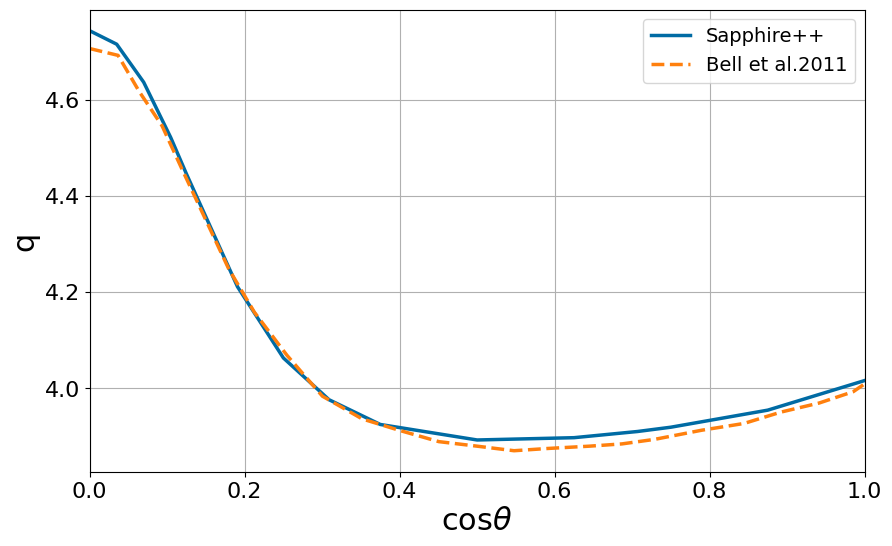}
  \caption{Magnitude of the spectral index $q$ at the shock for different shock obliquities, using the best fit to $f_{0}(0)\propto p^{-q}$.
    Here $\cos \theta = |B_x|/B$ is the angle between the shock normal and the magnetic field.
    $\nu =0.1 \omega_{\rm g}$. }
  \label{fig:4}
\end{figure}

\begin{figure}
  \centering{
       \,\,\,\includegraphics[width=0.94\linewidth]{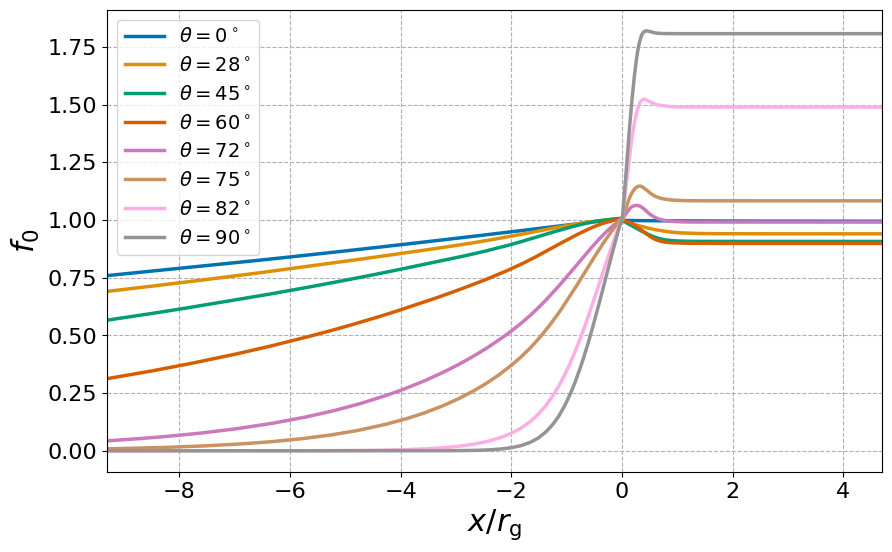}\\
    \hfill \includegraphics[width=0.9\linewidth]{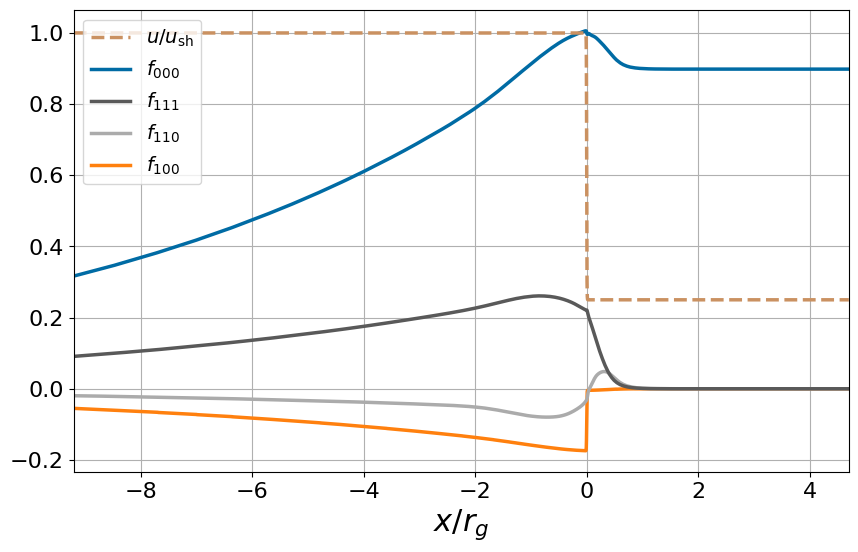}
  }
  \caption{Spatial profiles of particles with $p=20 mc$ at a shock with \(u_{\rm sh}= 0.1 c\) and Bohm-scattering rate \({\nu}/{\omega_{\rm g}}=0.1\) (Top) Plot of \(f_{0}\) for different angles of shock obliquity.
  (Bottom) Spatial profiles of coefficients \(f_{lms}\) for \(\theta = 60^\circ\).
  Shock profile is also shown for scale.
  All curves are normalized with respect to the value of $f_0$ at $x=0$.
  }
  \label{fig:5}
\end{figure}

As mentioned in the main text, \citet{Bell2011} previously studied steady-state solutions to the VFP equation at oblique shocks using a spherical harmonic decomposition.
To reduce computational cost, they made the replacement $\partial f_{lms}/\partial p = -q f_{lms}/p$.
In this way, the equations can be solved for a single value of $p$, to find the unknown quantity $q$.
This approach is expected to hold for Bohm scaling.
To test our code, we keep the full momentum dependence in our simulations but use matching physical conditions, with Bohm scaling for the scattering rate. 
\citet{Bell2011} consider a wide range of conditions.
We focus on a single shock velocity $u_{\rm sh} = 0.1 c$ and scattering rate $\nu = 0.1 \omega_{\rm g}$, and perform a scan of obliquity angles.
The steady-state solutions are well fit by a power-law, whose best fit parameters are compared to the matching curve from \citet{Bell2011} in FIG.~\ref{fig:4}, showing good agreement.
We checked that for the $\theta_{n}=60^\circ$ shocks, $l_{\rm max} =6$ was sufficient for convergence, and that at larger angles we used $l_{\rm max} =11$ (see also \cite{Bell2011}).

The spatial profiles of $f_{0}$ in the vicinity of the shock as a function of angle $\theta$ are shown in the upper panel of FIG.~\ref{fig:5}, while the leading first order ($l=1$) anisotropies for the $60^\circ$ case are shown in the lower panel.
For completeness, we note that the first few terms in the expansion can be expressed with our normalization of the real spherical harmonics as
\begin{align*}
  f(t,\mathbf{x}, \mathbf{p}) & =  \frac{f_{000}}{\sqrt{4 \pi}} + \sqrt{\frac{3}{4 \pi}} \left[f_{100}\frac{p_x}{p} -f_{110}\frac{p_y}{p} -f_{111}\frac{p_z}{p} \right]
\end{align*}
such that we can identify the particle current $\mathbf{j}$ as
\begin{equation*}
  \mathbf{j} = \int \mathbf{v} f d^3 p = \sqrt{\frac{4 \pi}{3}}\int v p^2
  \left( \begin{array}{c}
      f_{100} \\ -f_{110} \\ -f_{111}
    \end{array}\right) dp \, .
\end{equation*}
Thus, the $l=1$ coefficients have a simple interpretation in terms of differential particle fluxes.
Far from the shock, their solution can be found in the diffusion approximation \cite[e.g.][]{Reville2013}
\begin{align*}
  f_{000} \propto \exp\left(\int \, dx\,
  \frac{3 \eta}{r_{\rm g}} \frac{u_{\rm sh}}{v}\frac{\eta^2 +1}{\eta^2 + \cos^2 \theta_n}
  \right)
\end{align*}
and
\begin{align*}
  \left( \begin{array}{c}
             f_{100} \\ f_{110} \\ f_{111}
           \end{array}\right)
  = - \sqrt{3} \frac{u_{\rm sh}}{v}  \left( \begin{array}{c}
                                                \eta^2+\cos^2 \theta_n \\ \eta\sin\theta_n \\ -\cos\theta_n\,\sin\theta_n
                                              \end{array}\right)  \frac{f_{000}}{\eta^2+\cos^2 \theta_n}
\end{align*}
This can be seen in FIG.~\ref{fig:5}.
With the upstream $\mathbf{B}$ in the $x-z$ plane, the in-plane upstream flux is anti-aligned with the magnetic field, as particle diffuse against the flow i.e. $j_x\propto f_{100} < 0$ and $j_z\propto -f_{111} < 0$.
The out of plane anisotropic component, $f_{110}$ is due in part to cross field diffusion and the effect of the misaligned large-scale upstream particle gradient with respect to the magnetic field direction, i.e a gradient current.
We see that $j_y$ switches from positive to negative, consistent with the expected direction of $\nabla B$ drift as particles traverse the shock.

\begin{figure}
  \centering
  \includegraphics[width=0.9\linewidth]{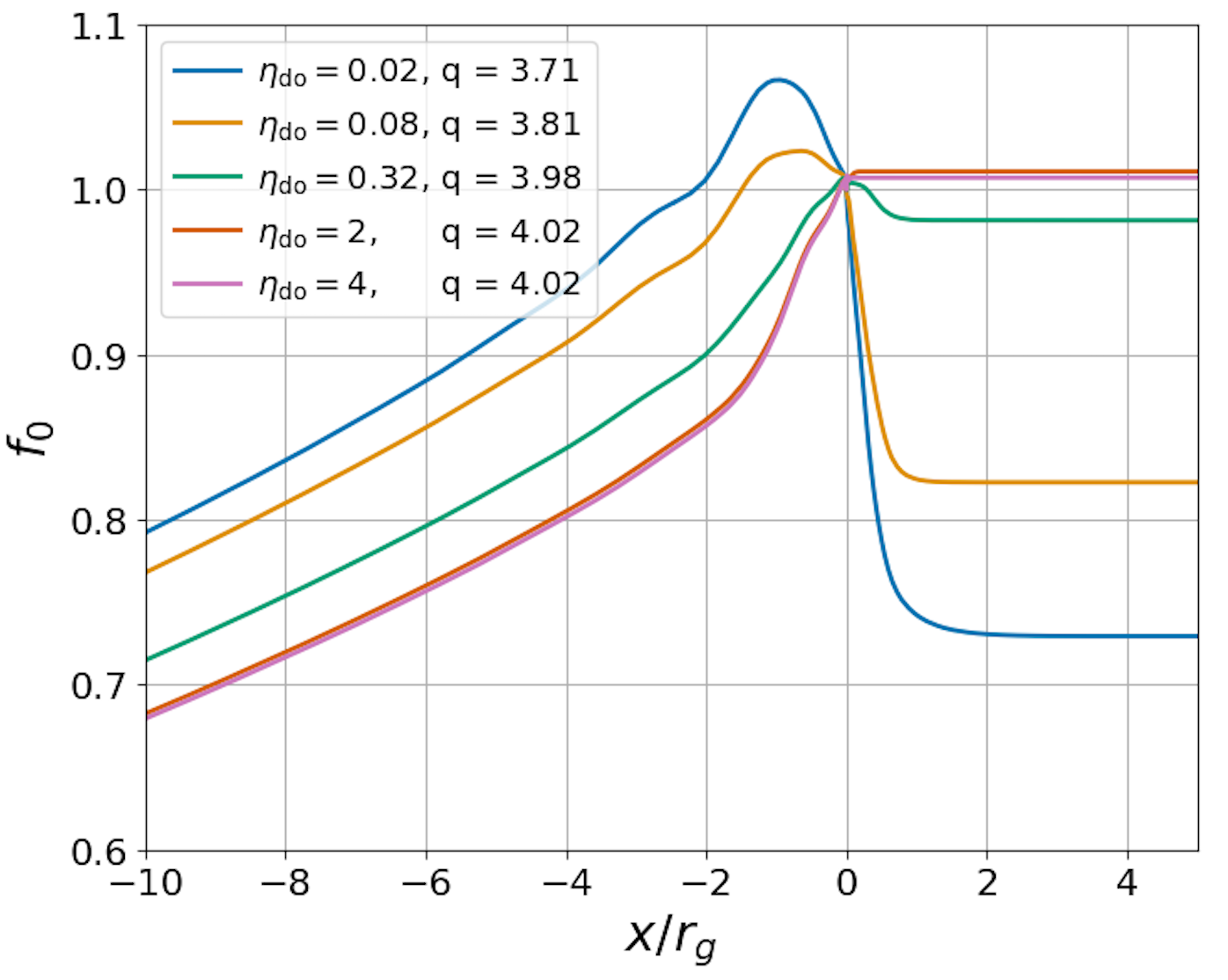}
  \caption{Spatial profiles for different scattering rates in the downstream half plane for $u_{\rm sh} = 0.1 c$ and $\theta_n=60^\circ$.
    Upstream scattering rate fixed to $\eta_{\rm up} = 0.02$.
    In all cases $p=20 mc$.
  }
  \label{fig:6}
\end{figure}

Finally, we note the dependence of the jump in $f_{000}$ at the shock on the scattering rate.
The asymptotic state is reached at a distance $\approx r_{\rm g}$ downstream of the shock.
In FIG.~\ref{fig:6} we consider weak upstream scattering ($\eta_{\rm up}=0.02$) but different scattering rates in the downstream for $\theta_n = 60^\circ$ and $u_{\rm sh}=0.1 c$.
It is found that as the scattering rate increases, the jump diminishes, and the spectral index approaches $4$.
For weak scattering upstream and down, a strong shock-reflected component is visible in the immediate upstream for subluminal shocks.

\bsp	
\label{lastpage}
\end{document}